\documentclass{article}

\usepackage{arxiv}

\usepackage[utf8]{inputenc} 
\usepackage[T1]{fontenc}    
\usepackage{hyperref}       
\usepackage{url}            
\usepackage{booktabs}       
\usepackage{amsmath,amssymb,amsfonts}       
\usepackage{nicefrac}       
\usepackage{microtype}      
\usepackage{lipsum}		
\usepackage{graphicx}
\usepackage{natbib}
\usepackage{doi}
\usepackage{caption}
\usepackage{subcaption}

\title{Competition and Investment Model of Wealth Distribution}

\author{ \href{https://orcid.org/0009-0004-3494-7339}{\includegraphics[scale=0.06]{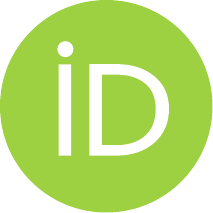}\hspace{1mm}Yuri Ono} \\
	Graduate School of Sociology\\
	Kwansei Gakuin University\\
	1-1-155 Uegahara Ichiban-cho, Nishinomiya, Hyogo 662-8501, Japan \\
	\texttt{ihe05244@kwansei.ac.jp} \\
	\And
	\href{https://orcid.org/0000-0001-5301-9838}{\includegraphics[scale=0.06]{orcid.pdf}\hspace{1mm}Atsushi Ishida} \\
	School of Sociology\\
	Kwansei Gakuin University\\
	1-1-155 Uegahara Ichiban-cho, Nishinomiya, Hyogo 662-8501, Japan \\
	\texttt{aishida@kwansei.ac.jp} \\
}

\hypersetup{
pdftitle={A template for the arxiv style},
pdfsubject={q-bio.NC, q-bio.QM},
pdfauthor={David S.~Hippocampus, Elias D.~Striatum},
pdfkeywords={First keyword, Second keyword, More},
}

\begin{document}
\maketitle

\begin{abstract}
	Explaining empirically observed wealth and income distributions, featuring power-law tails alongside gamma or log-normal bulk shapes, challenges models that focus on either pairwise competition or individual investment mechanisms. This study proposes and analyzes a unified model that integrates pairwise competition and individual investment via an adjustable parameter, $\alpha$. Numerical simulations are conducted to analyze the model's Gini coefficient and distributional shapes using the complementary cumulative distribution function and goodness-of-fit tests. Results show that the model captures a systematic transition in the bulk distribution from gamma like (low $\alpha$) to log-normal like (high $\alpha$). Additionally, intermediate levels of mechanism mixing can reduce inequality compared with the original mechanisms. However, it is difficult to distinguish heavy tails consistent with power-laws from log-normal tails. These findings highlight the importance of considering the interaction between different economic mechanisms but suggest that accurately replicating the empirical power-law tail requires more than the simple combination investigated.
\end{abstract}

\keywords{wealth distribution, gamma distribution, log-normal distribution, power-law}

\section{Introduction}\label{sec1}
Wealth distribution has been widely investigated over the years, and the mathematical modeling of its underlying mechanisms is extremely beneficial. An early attempt to mathematically model such mechanisms based on two-body interactions was proposed by John Angle. Angle formalized a stochastic process of competition for wealth through an interacting particle system and used it to explain income distribution in the United States \citep{angle1986,Angle1992,AngleDerivingSizeDistribution1993,Angle1996,Angle2002,ANGLE2006388,Angle2012,AngleHowWinAcceptance2013,Angle2022}.

This formalization, which is known as the inequality process, models wealth distribution through pairwise, zero-sum interactions among agents. This model is based on the concept of ``robust losers,'' which reflects the notion that more productive workers are better protected \citep{Angle2022}. The model operationalizes this notion via a small $\omega$ (which denotes a proportion of wealth lost by a loser). A key finding is that the situation of small $\omega$, which represents agents sheltered in the competition for wealth, results in a stationary wealth distribution that is well approximated by the gamma distribution \citep{AngleDerivingSizeDistribution1993,Angle2002}. Whereas Angle argues that the distribution generated under different conditions (large $\omega$) can approximate a Pareto-like form \citep{ANGLE2006388}, the mechanism yielding a gamma distribution is arguably more central to Angle's original theoretical motivation.

However, the mechanism proposed by Angle focuses on zero-sum competition and captures only one aspect of reality. The role of wealth creation or destruction, particularly through investment, is recognized as an important factor in shaping these distributions. Modeling wealth creation or destruction typically involves multiplicative stochastic processes based on Gibrat’s law of proportional effects \citep{gibrat1931}, in which income changes are proportional to income itself. Hamada provided a significant micro-foundation for this perspective \citep{HamadaGenerativeModelIncome2003, HamadaGenerativeModelIncome2004, HiroshiHAMADA2016} and reinterpreted this approach by developing a simple model based on a repeated investment game, in which an individual’s income changes multiplicatively depending on investment success or failure. Hamada showed that this investment mechanism generates an income distribution that is well approximated by a log-normal distribution.

The Angle and Hamada models assume the different types of mechanisms that generate wealth distributions and yield the different distributional outcomes (gamma and log-normal). These theoretical models should be evaluated based on empirical wealth and income distributions. Numerous empirical studies across various countries and time periods have established key stylized facts about wealth and income distributions \citep{pareto1964cours, LevyNewEvidencePowerlaw1997, Abul-MagdWealthDistributionAncient2002, HegyiWealthDistributionPareto2007}. Notably, the upper tail (high-wealth range) consistently follows a power-law (Pareto distribution), whereas the bulk of the distribution (low- and middle-wealth range) is typically better described by exponential, gamma, or log-normal distributions in the United States \citep{DrăgulescuEvidenceExponentialDistribution2001,DrăgulescuExponentialPowerlawProbability2001,SilvaTemporalEvolutionThermal2005,ClementiPowerlawTailExponent2006c, BanerjeeUniversalPatternsInequality2010}, UK \citep{DrăgulescuExponentialPowerlawProbability2001, ClementiParetosLawIncome2005,CoelhoFamilynetworkModelWealth2005}, Australia \citep{BanerjeeStudyPersonalIncome2006}, Germany \citep{ClementiPowerlawTailExponent2006c}, Italy \citep{ClementiPowerLawTails2005}, Romania \citep{DerzsyIncomeDistributionPatterns2012}, Japan \citep{SoumaUniversalStructurePersonal2000}, and China \citep{XuEvidenceChineseIncome2017}.

Comparing these empirical stylized facts with the outcomes of the theoretical models discussed earlier reveals certain limitations of the existing approaches. First, although these models successfully explain certain sections of the distribution for the bulk ranges (typically characterized by gamma or log-normal distributions), they do not sufficiently explain the generation mechanism of the power-law distribution observed in the upper tail. Second, in actual economic activities, individuals are likely to engage in competition and investment simultaneously. However, many existing models consider only one of these mechanisms and fail to capture the complex processes in which both interact. Consequently, current approaches offer only a partial explanation of the process for generating actual wealth and income distributions. Moreover, they fail to provide a unified understanding of the overall observed distribution, particularly the structure exhibiting different shapes in the bulk and upper tail.

Hence, this study is performed to propose a unified model that incorporates both the core mechanisms of existing representative models: pairwise competition and individual investment. This model allows one to adjust the relative contribution of wealth redistribution via competition and wealth growth via investment, thus providing a framework for examining the interaction between these two fundamental processes.

Using this unified model, we aim to address the following key questions: First, what overall distributional patterns emerge from the simultaneous operation of competitive and investment mechanisms? Second, how does the shape of wealth and income distributions, particularly in the bulk, change systematically as the relative contribution of investment versus competition varies? Finally, to what extent can this unified model replicate the empirically observed power-law tail, and under what conditions might such a behavior emerge?

The remainder of this paper is organized as follows: Section \ref{sec2} provides the general definition of the unified model. Section \ref{sec3} describes the method used to analyze the unified model. Section \ref{sec4} presents an analysis of the model based on numerical simulations and goodness-of-fit tests. Section \ref{sec5} presents conclusions and future research directions.

\section{Model Definition}\label{sec2}

This section defines the unified model used in our analysis. We first describe the two component models, i.e., Angle's inequality process representing pairwise competition and Hamada's repeated investment game representing individual investment, and then introduce the unified model that integrates them.

The models defined in this section are conceptualized as systems that exhibit the following general characteristics: This model includes numerous agents, which remain constant. Each agent $i$ possesses wealth\footnote{Regarding the terminology for the variable $x_{i(t-1)}$ representing the quantity held by agent $i$ at time $t-1$, previous studies used different terms: \cite{HiroshiHAMADA2016} used ``capital,'' whereas \cite{Angle2002} used ``wealth.'' However, both modeled $x_{i}$ as a stock concept. We herein use the term ``wealth'' consistently to refer to $x_{i}$.} $x_{i(t-1)}$ at time $t-1$, which evolves through stochastic transition rules. We can represent the model’s transition rule as a $2 \times 2$ matrix $\mathbf{M}$ as follows:
\begin{align}
    \begin{pmatrix}
    x_{it} \\ x_{jt}
    \end{pmatrix} &= \mathbf{M}
    \begin{pmatrix}
    x_{i(t-1)} \\ x_{j(t-1)}
    \end{pmatrix}.
\end{align}
The processes involve either pairwise interactions between agents or individual investments.

The transition rule for two interacting agents $i$ and $j$ in Angle's inequality process \citep{AngleDerivingSizeDistribution1993} can be represented by the matrix $\mathbf{M_C}$ as follows:
\begin{align}
      \mathbf{x_{t}} &= \mathbf{M_C} \mathbf{x_{t-1}},\\
\begin{pmatrix}
      x_{it} \\ x_{jt}
    \end{pmatrix} &= \begin{pmatrix}
      1 - \omega_{C}(1-C_{t}) & \omega_{C} C_{t} \\
      \omega_{C}(1- C_{t}) & 1 - \omega_{C} C_{t}
      \end{pmatrix} \begin{pmatrix}
        x_{i(t-1)} \\ x_{j(t-1)}
      \end{pmatrix},
\end{align}
where $x_{it}$ is the agent $i$'s wealth after encountering agent $j$. Wealth must remain non-negative, i.e., $x_{it} \geq 0 $. Here, $ C_{t} \in \{0,1\}$ is a random Bernoulli variable of value $1$ or $0$ ($1 = \text{win}; 0 = \text{lose}$), with fixed probability $p=1/2$ for a win in the inequality process. $\omega_{C}$ represents a proportion of wealth lost by the loser in the inequality process, and its range is $0 \leq \omega_{C} \leq 1$.

We consider two versions of the repeated investment game based on \cite{HamadaGenerativeModelIncome2003,HamadaGenerativeModelIncome2004,HAMADA2023}.
The first is the wealth-maintaining version, in which wealth is preserved even after investment failure. This version can be interpreted as representing safe asset building, in which wealth accumulates through investment success without the risk of loss. The second is the wealth-loss version, in which wealth decreases upon failure. This version can be interpreted as representing risky asset building, in which agents encounter potential losses and gains. Both types of repeated investment games yield the log-normal distribution. However, each investment game involves a distinct process. We focus on the wealth-maintaining version to investigate the effect of adding a basic wealth-creation process to a zero-sum competition\footnote{Applying the analytical procedures described in this paper to the wealth-loss version yielded qualitatively consistent results for the inequality patterns and distributional shape characteristics (bulk and tail fits). However, the mean wealth trends were distinct, with the mean remaining primarily constant with respect to $\alpha$ in the loss version.}.

The transition rule for the repeated investment game can be represented by the matrix $\mathbf{M_{I}}$ as follows:
\begin{align}
      \mathbf{x_{t}} &= \mathbf{M_{I}} \mathbf{x_{t-1}},\\
\begin{pmatrix}
      x_{it} \\ x_{jt}
    \end{pmatrix} &= \begin{pmatrix}
      1 + \omega_{I} I_{t} & 0 \\
      0 & 1 + \omega_{I}(1 - I_{t})
      \end{pmatrix} \begin{pmatrix}
        x_{i(t-1)} \\ x_{j(t-1)}
      \end{pmatrix},
\end{align}
where $I_{t} \in \{0,1\}$ is the Bernoulli variable with fixed probability $p=1/2$ for a win in the repeated investment game; and $\omega_{I}$ represents the proportion of wealth lost by the loser in the repeated investment game, which satisfies $0 \leq \omega_{I} \leq 1$.

We define the unified model of pairwise competition and individual investment as a linear convex combination parameterized by $\alpha$. The range $\alpha$ is $0 \leq \alpha \leq 1$. When $\alpha = 0$, the model reduces to the inequality process, whereas when $\alpha = 1$, the model becomes the repeated investment game. The parameter $\alpha$ quantifies the relative contribution of the two mechanisms, thus indicating whether wealth changes are primarily due to pairwise competition or individual investment.

The transition rule in the unified model can thus be expressed as
\begin{align}
      \mathbf{x_{t}} &= \{ (1 - \alpha) \mathbf{M_C} + \alpha \mathbf{M_{I}} \}\mathbf{x_{t-1}},\\
\begin{pmatrix}
      x_{it} \\ x_{jt}
    \end{pmatrix} &= \begin{pmatrix}
      1 - \omega_{C}(1-\alpha)(1-C_{t}) + \omega_{I}\alpha I_{t} & \omega_{C}(1-\alpha) C_{t} \\
      \omega_{C}(1-\alpha)(1-C_{t}) & 1 - \omega_{C}(1-\alpha) C_{t} + \omega_{I}\alpha(1-I_{t})
      \end{pmatrix} \begin{pmatrix}
        x_{i(t-1)} \\ x_{j(t-1)}
      \end{pmatrix}.
\end{align}

For analysis clarity and simplicity, we introduce the following assumptions: First, although the outcomes of pairwise competition $C_{t}$ and individual investment $I_{t}$ can be correlated, we assume they are independent:
\begin{align}
    \Pr(C_{t}, I_{t}) = \Pr(C_{t})\Pr(I_{t}).
\end{align}
This simplification is based on the notion that these two processes are driven by different forms of luck: competitive success may be related to luck specific to an individual pairwise encounter, whereas investment success may depend more on broader market luck or factors external to the interaction.

Second, we assume a common parameter $\omega$ for both mechanisms, although they may differ in principle. This simplification is based on our primary focus in this study, which is to investigate the effects of the parameter $\alpha$ representing the integration degree. We interpret $\omega$ as a parameter indicating the overall rate or intensity of wealth fluctuation in the system. Thus, we set
\begin{align}
    \omega = \omega_{C} = \omega_{I}.
\end{align}
These assumptions are maintained throughout the paper.

\section{Method}\label{sec3}

This section details the method used to analyze the unified model proposed in Section \ref{sec2}. We describe the simulation procedures and analytical methods, including parameter setting, the inequality index employed, and goodness-of-fit tests. 

\subsection{Simulation Settings}
We simulate a system with $N = 1000$ agents, each initially having a wealth of $x_0 = 10$. The simulation proceeded as follows: At every time step $t$, a total of $500$ pairs were randomly formed and interacted according to the model rules, thus resulting in a winner and loser. Wealth was transferred from the loser to the winner for interacting pairs. This process was repeated for $T=100$ time steps, which constituted a single trial, and the statistics were recorded. The mean and standard deviation of these statistics were obtained by averaging over $100$ independent trials.

For the goodness-of-fit analysis, we generate $100$ independent sets of wealth distributions at $T = 100$ for each parameter pair ($\omega$, $\alpha$). To obtain a smooth, representative distribution, we averaged $100$ independent stationary distributions based on the rank order. Specifically, within each trial, we arranged the $N=1000$ wealth in the ascending order. Subsequently, for each rank, we calculated the mean wealth across the $100$ trials. Goodness-of-fit tests were performed on the smoothed distributions.

We employed the Gini coefficient as a metric for inequality because it uses the information throughout the distribution and is widely employed, thus rendering it generalizable. The Gini coefficient ranges from $0$ to $1$. A Gini coefficient that is approximately $0$ indicates a perfectly equal state in which every agent has equal wealth, whereas a Gini coefficient that is approximately $1$ indicates a perfectly unequal state in which all income is concentrated in the hands of one agent.

\begin{figure}[h]
    \centering
        \includegraphics[scale = 0.5]{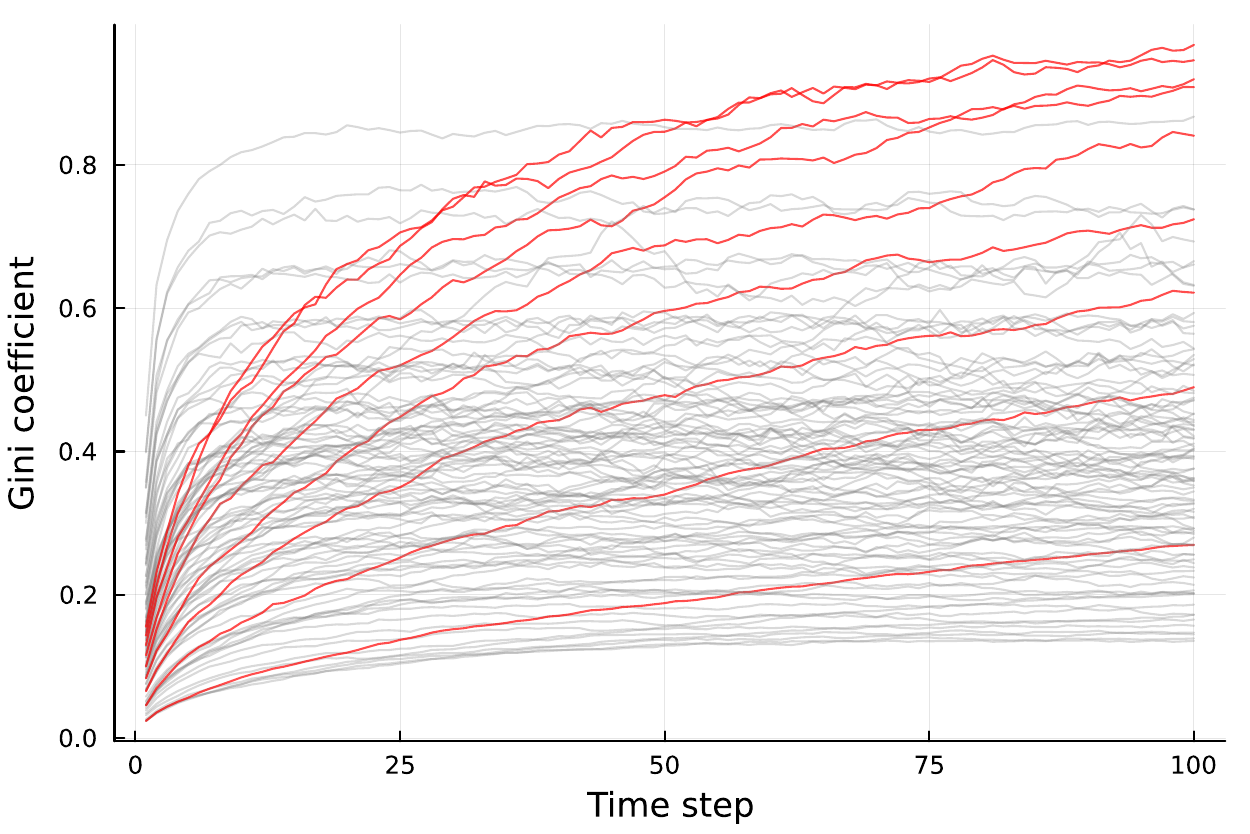}
        \caption{ Gini coefficient as a function of time $T$ for different values of $(\omega,\alpha)$, with case $\alpha=1.0$ highlighted}
        \label{fig: gini convergence with maintain}
\end{figure}

To validate the use of simulation results at time $T=100$ for our analysis, we examined the time evolution of the Gini coefficient based on \cite{CaonUnfairConsequencesEqual2007}. Figure \ref{fig: gini convergence with maintain} shows the time series for all the parameter sets. Except for the case $\alpha=1.0$, the Gini coefficient converged rapidly to a stable value after a short initial transient, thus indicating that the wealth distribution reached a statistically stable state well before $T=100$.

For $\alpha=1.0$, which corresponds to Hamada's repeated investment game, the mean and variance increased with simulation time $T$ \citep{HamadaGenerativeModelIncome2004}. Because the overall scale of the distribution continues to shift even as its shape converges, the Gini coefficient alone is insufficient for assessing stationarity in this case $(\alpha=1.0)$. However, the distribution-fitting results presented in Figure \ref{fig:CCDF_gamma_lnorm} suggest that the distributional form converged sufficiently by $T=100$. Based on these considerations, we regarded the simulation results at $T=100$ as the stationary distribution for the subsequent analysis in this study.

\subsection{Method of Goodness-of-fit Tests}

The primary goal of our analysis was to compare the distributions generated by the proposed model with the empirical characteristics of wealth and income distributions observed in real-world data. If the model captures the underlying mechanisms well, it should reproduce the key stylized facts. Accordingly, we examined the extent to which the unified model replicated these empirical properties. Specifically, we investigated whether a power-law distribution appropriately describes the upper tail of the generated distributions. For the bulk of the distribution, we assessed whether the gamma or log-normal distribution provided a better fit.

Notably, previous studies \citep{NewmanPowerLawsPareto2005,ClausetPowerLawDistributionsEmpirical2009} demonstrated that multiplicative stochastic processes, such as the investment component in our model, can occasionally generate distributions whose upper tails resemble a power-law but are in fact better described by a log-normal distribution. Therefore, a meticulous comparison is necessary.

To assess the goodness-of-fit for the upper tail, we first applied the approach proposed by \cite{ClausetPowerLawDistributionsEmpirical2009}. This involves estimating the lower bound of the power-law behavior, denoted as $x_{\min}$, and the shape parameter $a$ (Pareto exponent) by minimizing the Kolmogorov–Smirnov (KS) statistic between the data and the fitted power-law. Subsequently, we evaluated the plausibility of the power-law hypothesis using a goodness-of-fit test based on the KS statistic, which yielded a $p$-value. A $p$-value exceeding $0.1$ implies that the power-law model cannot be precluded.

Additionally, to directly compare the fit of the power-law against the log-normal distribution for the upper tail ($x \geq x_{\min}$) based on \cite{ClausetPowerLawDistributionsEmpirical2009}, we employed the Vuong test, which is a likelihood ratio test designed for non-nested model comparison \citep{VuongLikelihoodRatioTests1989}. The test statistic used was the normalized log-likelihood ratio, which asymptotically follows the standard normal distribution under the null hypothesis. Here, the null hypothesis is that the competing models—the power-law and log-normal distributions—are equivalent in terms of the Kullback-Leibler information criterion (KLIC) which measures the distance between the true distribution and a specified model. If the $p$-value is less than $0.1$, then the power-law distribution is favored over the log-normal. If the $p$-value exceeds $0.9$, then the log-normal distribution is favored over the power-law. Otherwise ($0.1\leq p\text{-value} \leq 0.9$), the models are considered statistically indistinguishable \citep{VuongLikelihoodRatioTests1989}.

Similarly, for the bulk of the distribution ($x < x_{\min}$), we compared the goodness-of-fit between the log-normal and gamma distributions using the Vuong test. In this comparison, observations where $x \geq x_{\min}$ were regarded as right censored at $x_{\min}$ to account for the possibility that the underlying process for these observations belongs to the bulk distribution. Here, the null hypothesis is that the competing models—the gamma and log-normal distributions—are equivalent in terms of the KLIC. If the $p$-value is less than $0.1$, then the log-normal distribution is favored over the gamma distribution. If the $p$-value exceeds $0.9$, the gamma distribution is favored over the log-normal distribution. Otherwise ($0.1\leq p\text{-value} \leq 0.9$), the two models are statistically indistinguishable.

Our primary goal in performing the fitting is to identify the most appropriate distributional form without obtaining precise parameter estimates. Therefore, data rescaling can be applied as required to facilitate the estimation process.

Regarding software solutions, the simulations in this study were performed using Julia 1.10.6 \citep{Julia-2017}, and subsequent data analyses were conducted using R 4.4.2 \citep{Rcoreteam}. The ``PoweRlaw'' \citep{Gillespie2015} package was used for power-law fitting and testing, and the ``fitdistrplus'' \citep{fitdistrplus} package was used to fit gamma and log-normal distributions with censored data and testing.

\section{Results}\label{sec4}

Based on the methodology outlined in Section \ref{sec3}, this section reports the numerical-simulation results obtained using the unified model. We first investigate the effect of parameters $\omega$ and $\alpha$ on the aggregate measures (the mean and Gini coefficient) and then evaluate the distributional shapes using both visual inspection (log-log complementary cumulative distribution function (CCDF) plots) and formal goodness-of-fit tests.

\subsection{Analysis of Mean and Inequality}

We present the simulation results, beginning with an analysis of the behavior of the aggregate measures for the wealth distribution, specifically the mean wealth and the Gini coefficient, under varying parameters ($\omega,\alpha$).

\begin{figure}[h]
    \centering
        \includegraphics[scale = 0.5]{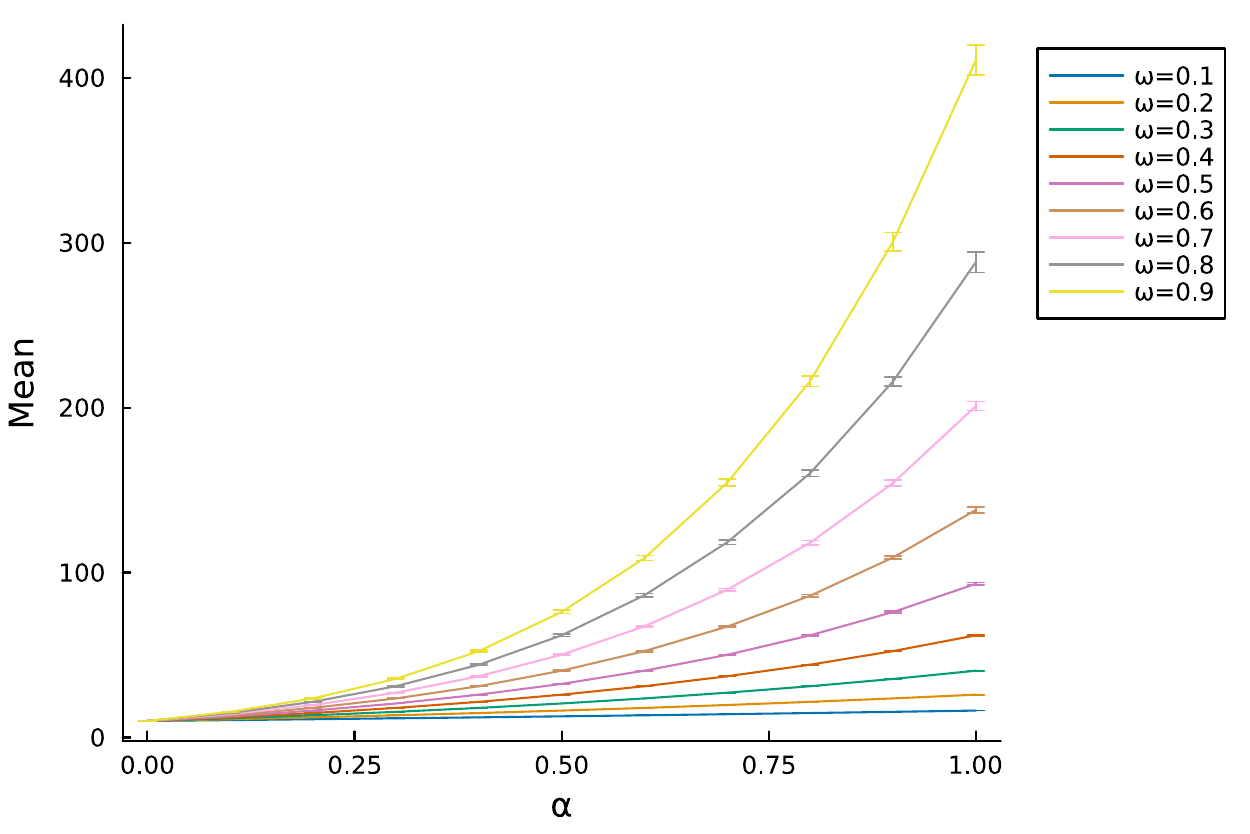}
        \caption{ Mean in unified model}
        \label{fig: mean with maintain}
\end{figure}

Figure \ref{fig: mean with maintain} shows the change in the average mean of the wealth distribution (averaged over $100$ independent trials) with the parameter $\alpha$ and the proportion of wealth lost by the loser, $\omega$, at $T=10$ under the unified model\footnote{Although presented for $T=10$ for visual clarity, the observed tendencies did not change substantially for larger $T$.}. The error bars represent the standard deviation across these $100$ trials. The mean increased with $\alpha$. Similarly, an increase in $\omega$ increased the mean. Moreover, the sensitivity of the mean to changes in $\omega$ became more pronounced as $\alpha$ increased.

\begin{figure}[h]
    \centering
        \includegraphics[scale = 0.5]{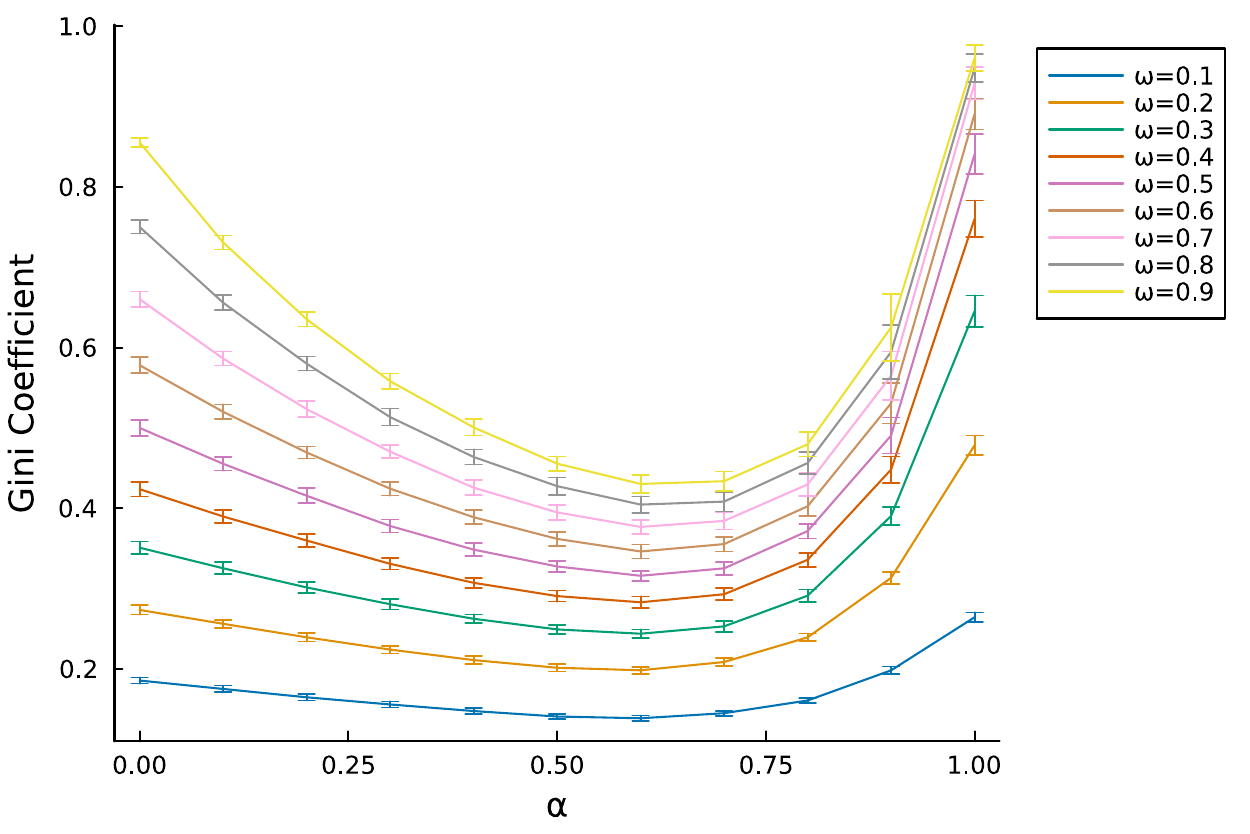}
        \caption{ Gini coefficient in unified model}
        \label{fig: gini with maintain}
\end{figure}

Figure \ref{fig: gini with maintain}  illustrates the change in the average Gini coefficient (averaged over 100 independent trials) with the parameters $\alpha$ and $\omega$ at $T = 100$ under the unified model. The error bars represent the standard deviation across these $100$ trials. The Gini coefficient increased with $\omega$ and exhibited a U-shaped trend. Compared with the boundary cases ($\alpha=0$ and $\alpha=1$), the Gini coefficient decreased at intermediate values of $\alpha$. This reduction was particularly pronounced at approximately $\alpha=0.6,0.7$.

\subsection{Shape of Wealth Distribution}

Following the analysis of the aggregate statistics, we examined the distributional shapes generated by the unified model in detail. First step, we visually inspected the distributions using CCDF plots on a log-log scale, which is useful for identifying potential power-law behavior in the tail.

\begin{figure}[htbp]
  \centering
  \begin{subfigure}[b]{0.48\textwidth}
    \includegraphics[width=\linewidth]{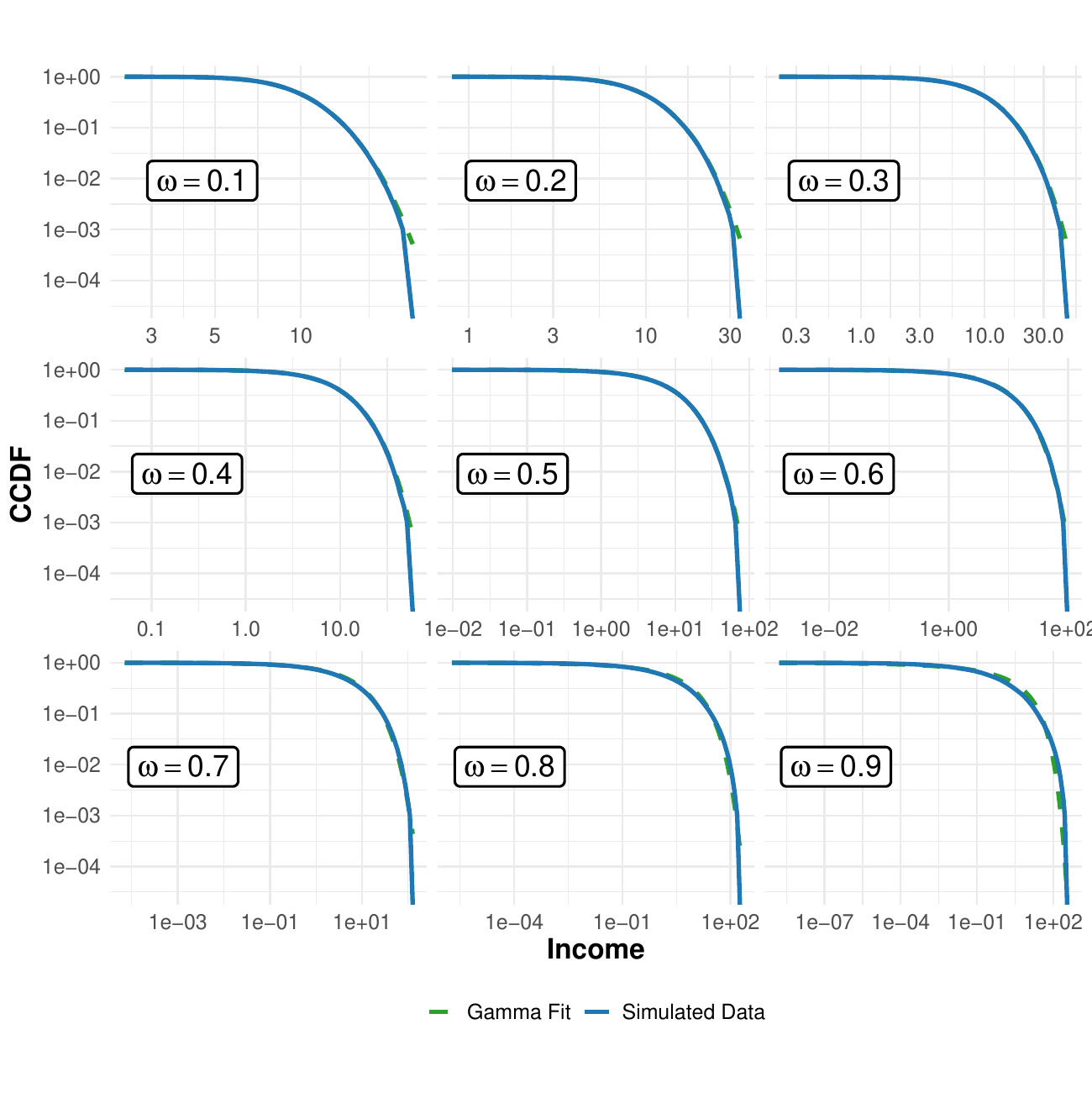}
    \caption{CCDF and gamma fits ($\alpha=0.0$)}
    \label{fig:CCDF_gamma}
  \end{subfigure}
  \hfill
  \begin{subfigure}[b]{0.48\textwidth}
    \includegraphics[width=\linewidth]{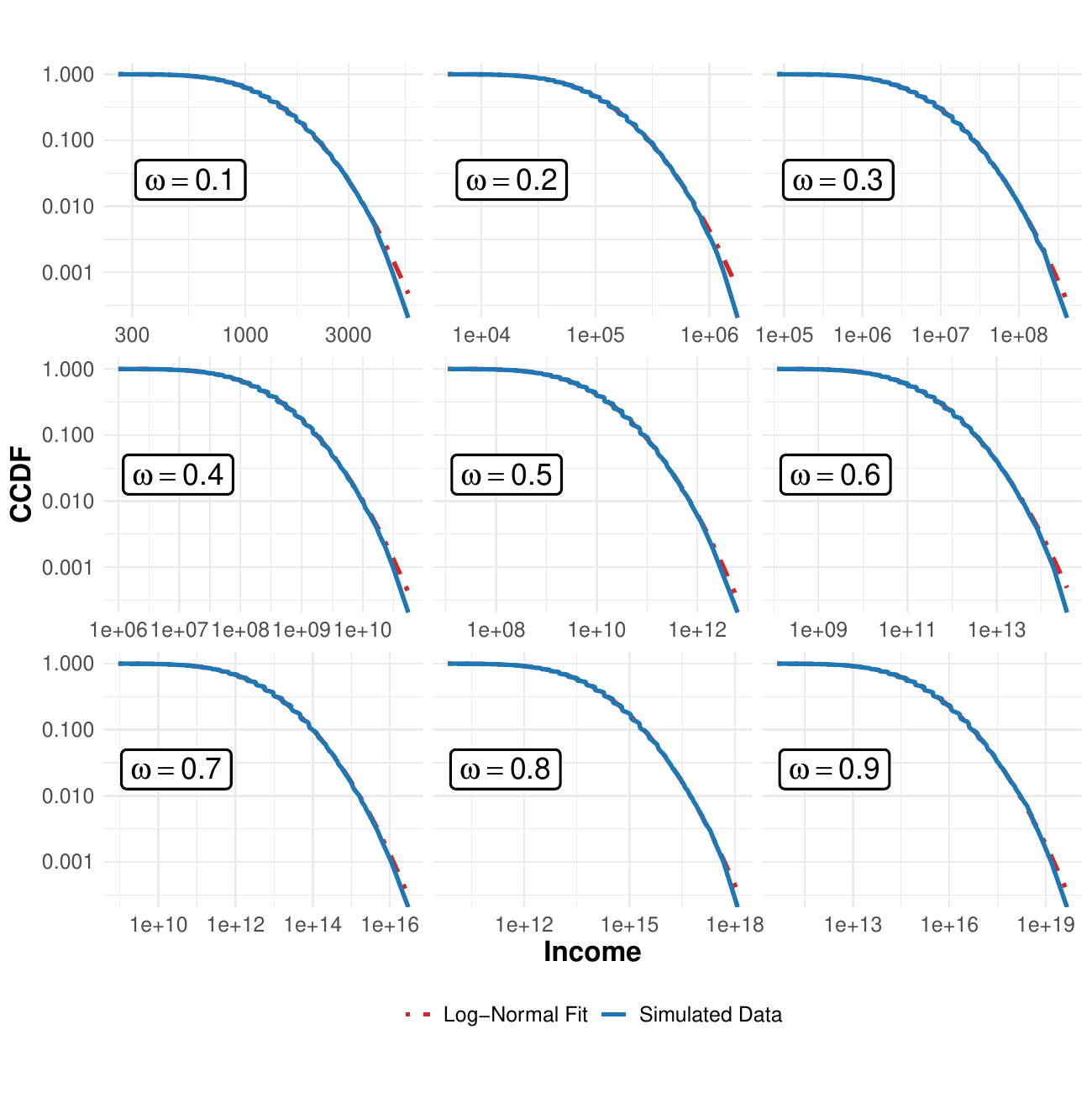}
    \caption{CCDF and log-normal fits ($\alpha=1.0$)}
    \label{fig:CCDF_lnorm}
  \end{subfigure}
  \caption{CCDF plots and corresponding theoretical fits ($\alpha=0.0$ and $1.0$)}
  \label{fig:CCDF_gamma_lnorm}
\end{figure}

Figure \ref{fig:CCDF_gamma_lnorm} presents log-log plots of the CCDF for the simulated data obtained using the unified model for $\alpha=0.0$ and $\alpha=1.0$. In each plot, the empirical CCDF is shown alongside the theoretical CCDF curve that is expected to fit the data (gamma distribution for $\alpha=0.0$; log-normal distribution for $\alpha=1.0$). This figure confirmed that our simulation reproduced the expected theoretical distributions at these parameter boundaries.

\begin{figure}[htbp]
  \centering

  \begin{subfigure}[b]{0.45\textwidth}
    \includegraphics[width=\linewidth]{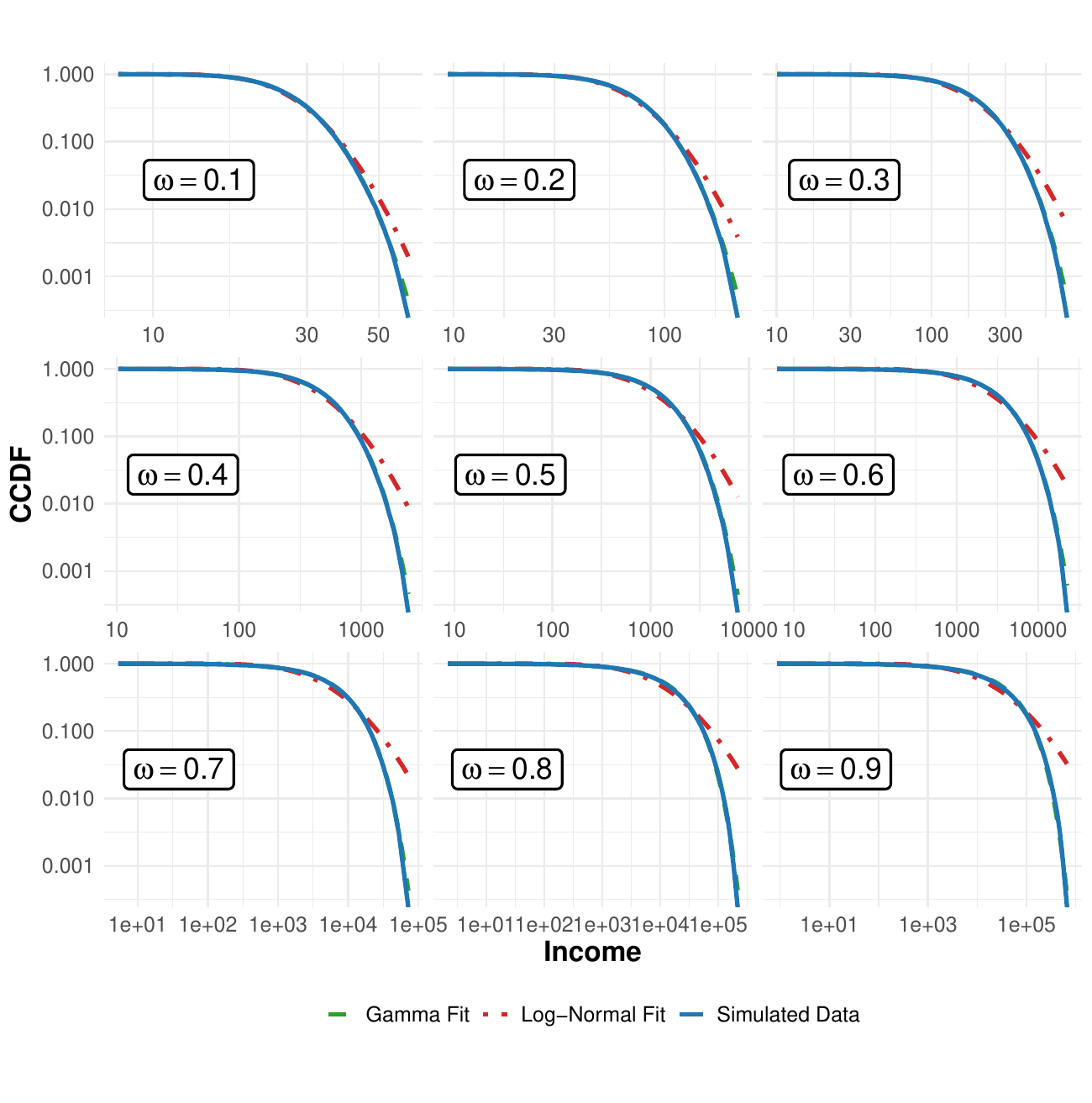}
    \caption{CCDF and fits for $\alpha=0.2$}
    \label{fig:fig1}
  \end{subfigure}
  \hfill
  \begin{subfigure}[b]{0.45\textwidth}
    \includegraphics[width=\linewidth]{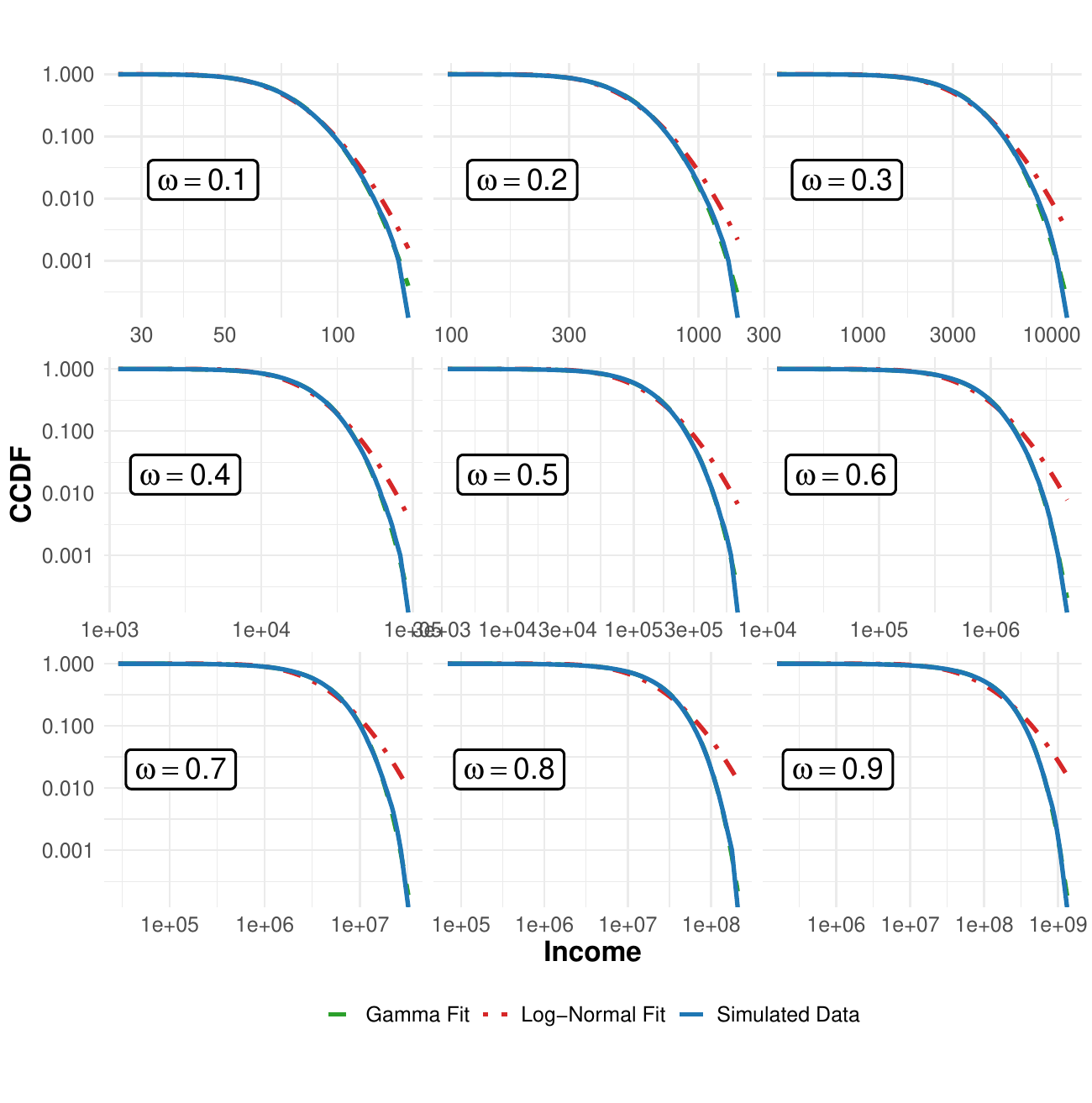}
    \caption{CCDF and their fits for $\alpha=0.4$}
    \label{fig:fig2}
  \end{subfigure}

  \vspace{1em}

  \begin{subfigure}[b]{0.45\textwidth}
    \includegraphics[width=\linewidth]{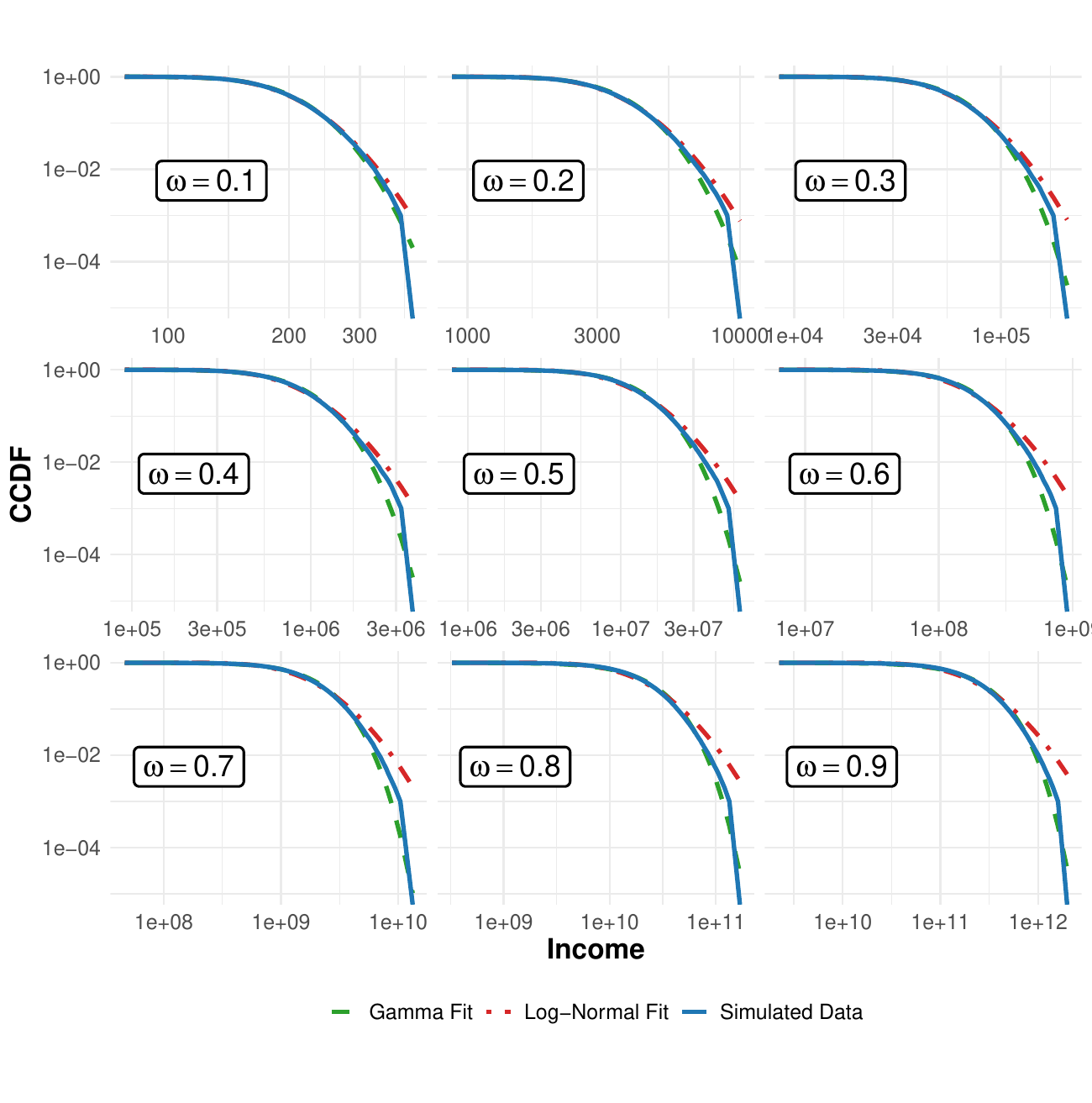}
    \caption{CCDF and fits for $\alpha=0.6$}
    \label{fig:fig3}
  \end{subfigure}
  \hfill
  \begin{subfigure}[b]{0.45\textwidth}
    \includegraphics[width=\linewidth]{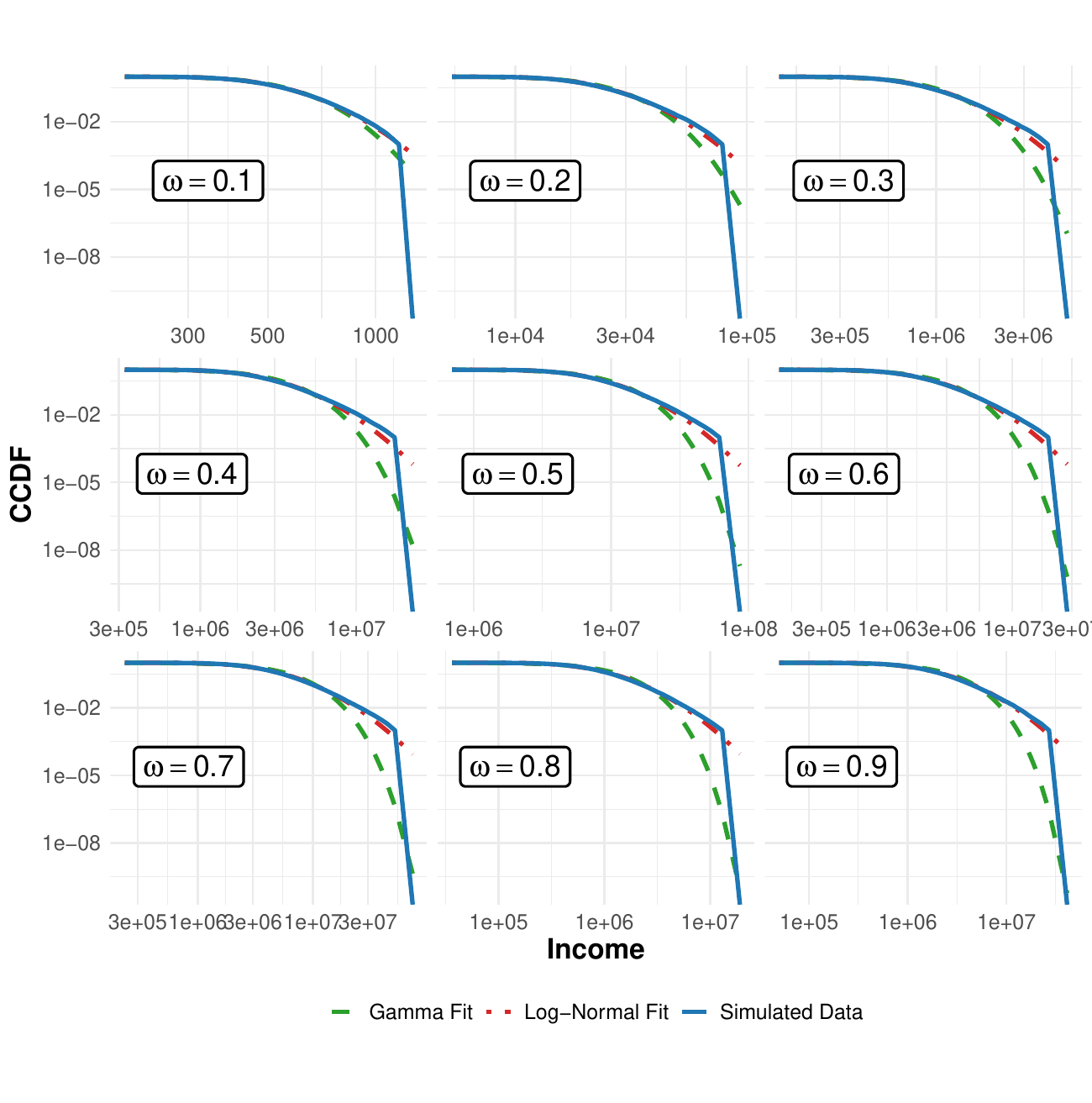}
    \caption{CCDF and fits for $\alpha=0.8$}
    \label{fig:fig4}
  \end{subfigure}

  \caption{Log-log CCDF plots for intermediate $\alpha$ values with gamma and log-normal fits}
  \label{fig:fourpdfs}
\end{figure}

After confirming the model's behavior at the boundaries ($\alpha=0.0$ and $\alpha=1.0$), we visually inspected the distribution-shape evolution for intermediate $\alpha$ values, specifically focusing on $\alpha \in\{0.2,0.4,0.6,0.8\}$. Figure \ref{fig:fourpdfs} presents log-log CCDF plots for the simulated data obtained from the unified model for these intermediate parameters. Theoretical CCDF curves for the fitted gamma and log-normal distributions are shown for comparison. Visual inspection of Figure \ref{fig:fourpdfs} suggests a systematic transition in the distribution shape as $\alpha$ increases. For small $\alpha$ values of $0.2$ and $0.4$, the distribution resembled the gamma fit more closely than the log-normal fit. However, this agreement with the gamma distribution weakened for larger $\alpha$ values of $0.6$ and $0.8$. Conversely, the log-normal fit matched the bulk increasingly well at higher $\alpha$, whereas it consistently failed to include the upper tail of the empirical distribution. This discrepancy in the tail became more pronounced as the mixing parameter $\alpha$ increased.

These observations suggest that the generated distribution, particularly for larger $\alpha$, exhibits distinct behaviors in the bulk vs. the tail, instead of conforming to a single, simple theoretical distribution such as the gamma or log-normal distribution across its entire range. Furthermore, the relatively straight appearance of the empirical CCDF in the tail region (the top 0.1\%) on these log-log plots is consistent with power-law behavior. To rigorously evaluate the distributional characteristics suggested by this visual approach, we performed quantitative goodness-of-fit tests.

\begin{figure}[h]
    \centering
        \includegraphics[scale = 0.5]{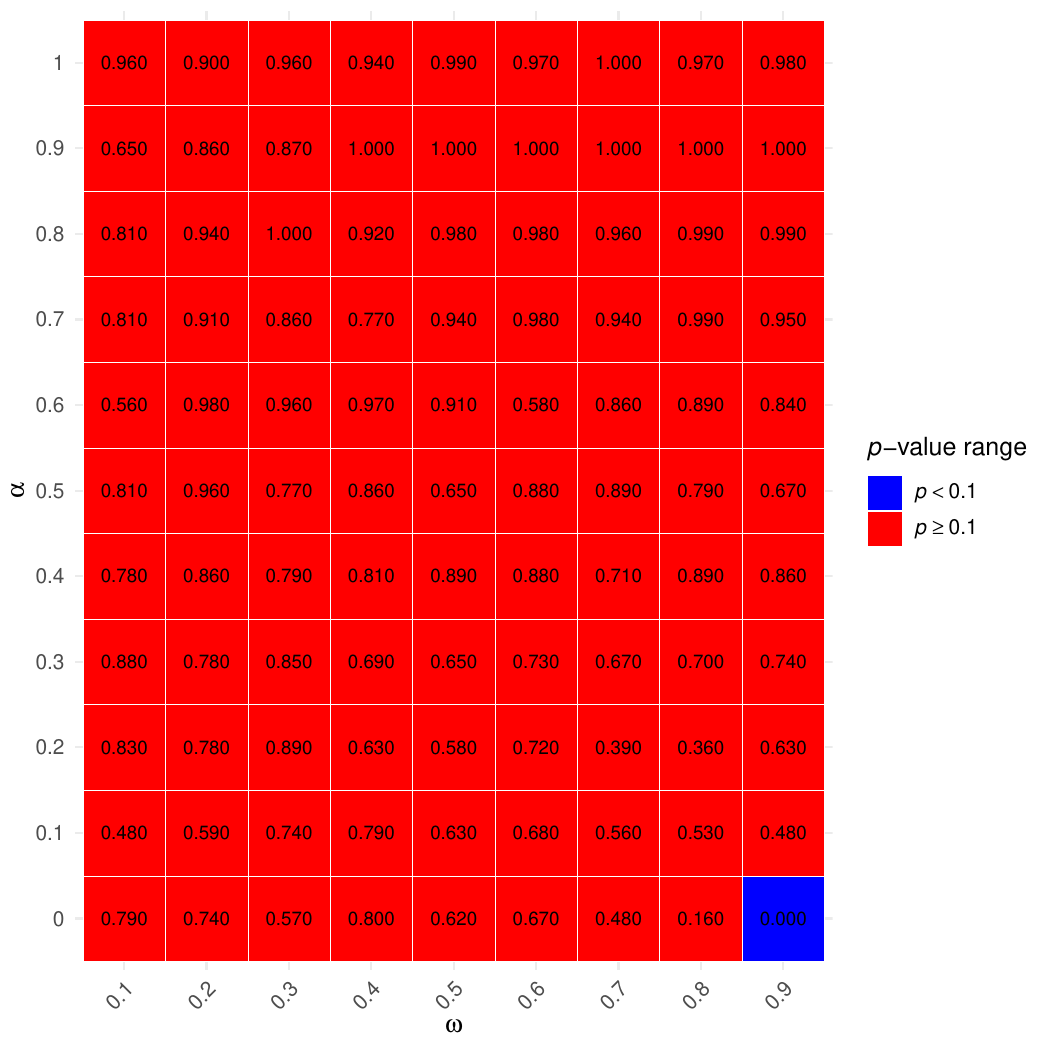}
        \caption{Heatmap of $p$-values from goodness-of-fit tests for power-law distribution in upper tail ($x\geq x_{\min}$) for each $\alpha$ and $\omega$ in unified model. Blue represents rejection ($p\text{-value} < 0.1$) and red represents non-rejection ($p\text{-value} \geq 0.1$) of power-law hypothesis.}
        \label{fig: p_value_high_main}
\end{figure}

Figure \ref{fig: p_value_high_main} shows a heatmap of $p$-values generated from the power-law goodness-of-fit test for the upper tail in the unified model. It is color coded based on whether the values exceed the threshold value of $0.1$. If the $p$-value is less than $0.1$, the null hypothesis that the distribution is generated based on the power-law is rejected \citep{ClausetPowerLawDistributionsEmpirical2009}. Consequently, the hypothesis of a generation process based on the power-law is not rejected\footnote{This non-rejection of the power-law hypothesis occurred at the boundaries $\alpha=0.0$ and $\alpha=1.0$. When considering the entire distribution shape as visualized in Figure \ref{fig:CCDF_gamma_lnorm}, the distribution at $\alpha=0.0$ and $\alpha=1.0$ is well approximated by the gamma and log-normal distributions, respectively. This suggests that whereas the tail alone might exhibit features that are relatively consistent with a power-law according to this test, the overall distributions in these boundary cases are better characterized by gamma and log-normal distributions.}, except for $(\omega,\alpha)=(0.9,0)$.

We directly compared two distributions against one another: for the upper tail ($x\geq x_{\min}$), we compared the power-law distribution against the log-normal distribution; whereas for the bulk ($x < x_{\min}$), we compared the gamma distribution against the log-normal distribution.
 
\begin{figure}[h]
    \centering
        \includegraphics[scale = 0.5]{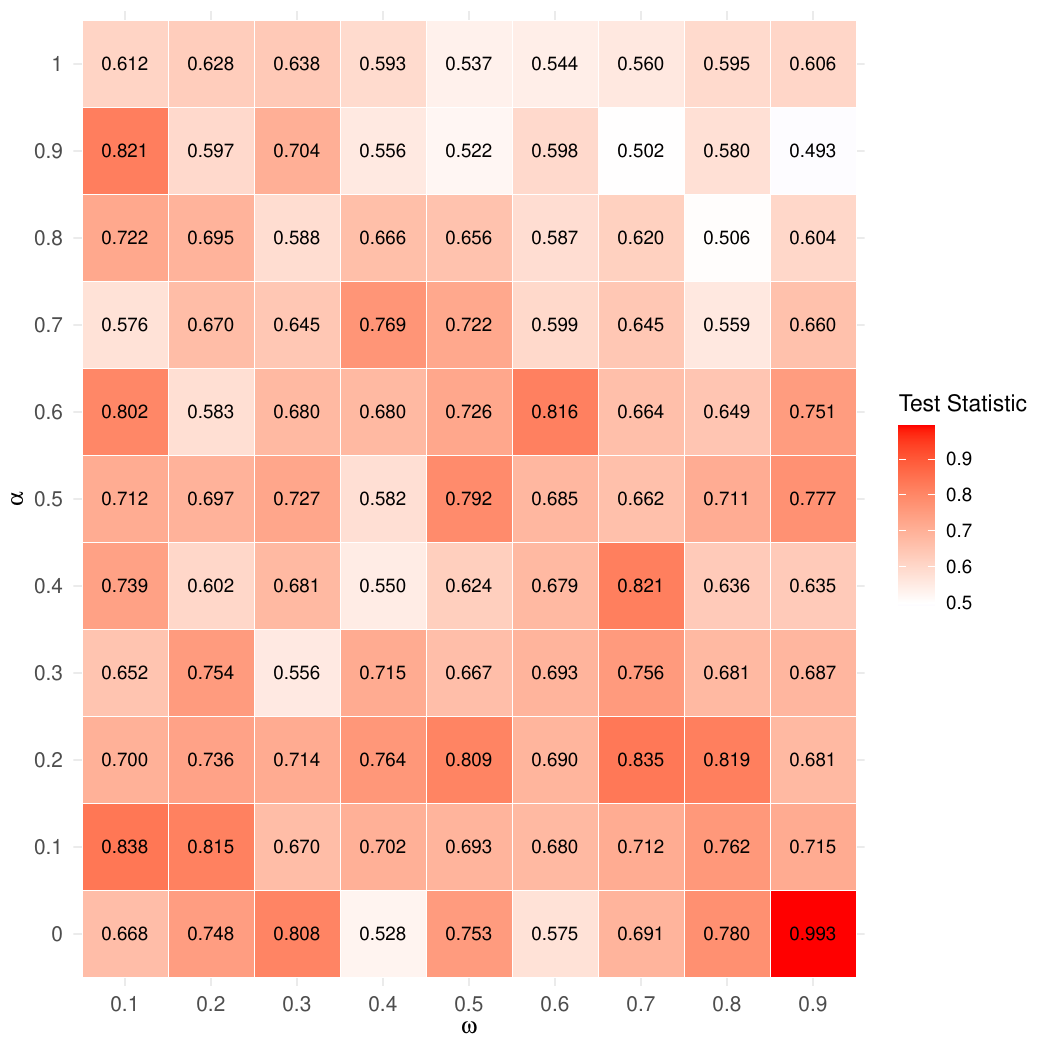}
        \caption{ Vuong test $p$-values comparing power-law vs. log-normal distributions for upper tail ($x\geq x_{\min}$) across different combinations of parameters ($\omega$, $\alpha$) in unified model. $p\text{-value} < 0.1$ favors power-law, whereas $p\text{-value} > 0.9$ favors log-normal; intermediate values ($0.1\leq p\text{-value} \leq0.9$) indicate that the models are indistinguishable based on KLIC.}
        \label{fig: vuong_results_high_main}
\end{figure}

Figure \ref{fig: vuong_results_high_main} shows the $p$-value based on Vuong test of distribution models (log-normal and power-law models) that better fit the upper tail ($x\geq x_{\min}$) for each parameter set ($\omega$, $\alpha$) in the unified model. No set can discriminate between the two competing models under all parameters except the parameter $(\omega, \alpha) = (0.9, 0)$, i.e., where the power-law fit is already rejected ($p\text{-value}< 0.1$).

\begin{figure}[h]
    \centering
        \includegraphics[scale = 0.5]{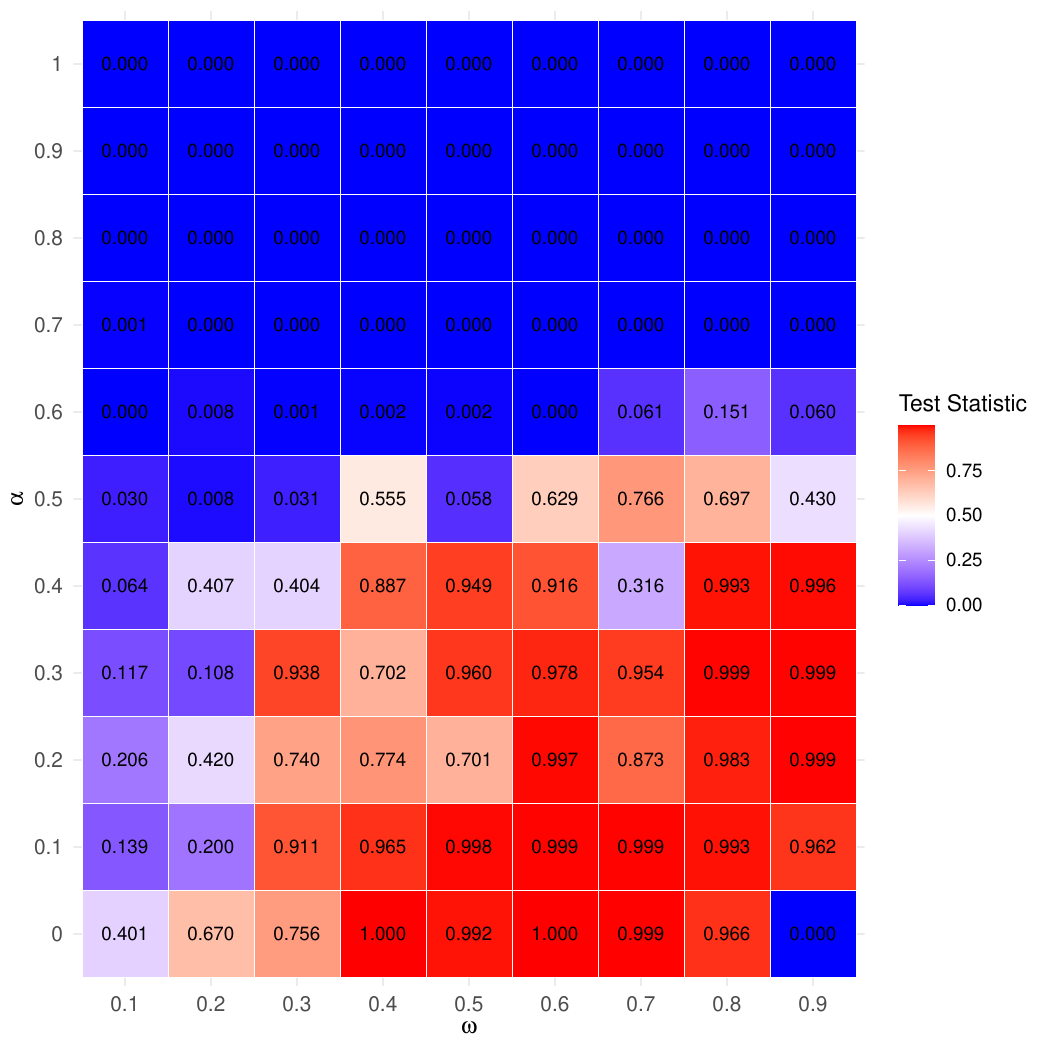}
        \caption{Vuong test $p$-values comparing gamma vs. log-normal distributions for bulk ($x<x_{\min}$, right censored at $x_{\min}$) across different combinations of parameters ($\omega$, $\alpha$) in unified model. $p\text{-value} < 0.1$ favors log-normal, whereas $p\text{-value} > 0.9$ favors gamma; intermediate values ($0.1\leq p\text{-value}\leq0.9$) indicate models are indistinguishable based on KLIC.}
        \label{fig: vuong_results_low_main}
\end{figure}

Next, we examine the distribution shape of the bulk ($x<x_{\min}$). Figure \ref{fig: vuong_results_low_main} shows the $p$-value based on Vuong test of distribution models (gamma and log-normal models) that better fit the bulk ($x<x_{\min}$) for each parameter set ($\omega$, $\alpha$) in the unified model. Consistent with the expectation from the models, the Vuong test results primarily indicate that the gamma distribution provides a significantly better fit for smaller values of $\alpha$, whereas the log-normal distribution is favored for larger values of $\alpha$, particularly in the unified model.

\section{Conclusion and Future Endeavors}\label{sec5}

Existing mathematical models explaining the distribution of wealth and income have limitations, primarily because they focus on either pairwise competition or individual investment mechanisms and typically fail to adequately reproduce empirically observed distributions, particularly the power-law tail observed in the high-wealth range. We address this challenge by proposing and analyzing the unified model, which integrates pairwise competition and individual investment mechanisms via an adjustable parameter $\alpha$. This model combines the core mechanisms of existing representative approaches, i.e., the pairwise competition model, whose stationary distribution follows a gamma distribution, and the individual investment model, whose stationary distribution follows a log-normal distribution. This model allows us to adjust the relative contributions of both mechanisms, thereby providing a framework for examining the manner by which their interaction shapes the overall wealth distribution.

Our simulation analysis revealed several key findings. First, for the unified model, intermediate values of $\alpha$ showed a lower Gini coefficient compared with the boundary cases ($\alpha=0$ and $\alpha=1$) when the proportion of wealth lost by the loser $\omega$ was high. Second, the proposed model captured the systematic transition of the bulk distribution from gamma like (at small $\alpha$) to log-normal like (at large $\alpha$) with $\alpha$. Third, regarding the upper tail, although the simulated data from the proposed model were plausibly drawn from a power-law distribution under certain conditions, distinguishing it definitively from a log-normal distribution was challenging.

These findings indicate that combining competition and investment mechanisms in the unified model provides important insights. The proposed model revealed unexpected behaviors that not revealed when examining the individual processes alone. The observation that inequality can decrease at intermediate $\alpha$ values under high $\omega$ is noteworthy, particularly because both competition and investment, considered separately, tend to increase inequality, which is consistent with their inherently diffusive nature \citep{GreenbergTwentyfiveYearsRandom2024}. An intuitive explanation is that competition redistributes gains from investment, thus offering reinvestment opportunities to the poorest agents. This prevents the extreme concentration typically driven by competition or investment alone and supports broader participation, thereby resulting in the observed reduction in inequality at intermediate $\alpha$. This highlights the benefit of examining the interaction between these mechanisms. Furthermore, whereas the original Angle and Hamada models typically yield distributions described by a single function (gamma or log-normal), their combination results in scenarios where the upper tail shows behavior that is potentially consistent with a power-law. Although definitively distinguishing this tail from a log-normal distribution is challenging, it can highlight the complexity introduced by unifying the two mechanisms. In fact, the appearance of such heavy tails, which potentially exhibit power-law characteristics despite originating from the combination of gamma and log-normal generating processes, can be interpreted as an emergent property of this unified model.

Additionally, the model demonstrated the ability to represent diverse distributional patterns observed empirically based on previous studies. It captured a smooth transition in the bulk of the distribution, from gamma like (low $\alpha$) to log-normal like (high $\alpha$), by adjusting the parameter $\alpha$. This finding suggests the synergistic effect of competition and investment may be essential for understanding the distribution patterns observed in the real world. We interpret this result as indicating that the observed preference for either gamma or log-normal fits in empirical studies may be driven by the relative strength ($\alpha$) of the two underlying processes operating in different societies or time periods.


Our findings regarding the upper tail suggest partial success but also reveal some critical limitations. First, the simple combination of competition and investment mechanisms might not be sufficient to fully replicate the clear power-law pattern observed in the upper tail. A more accurate matching of the upper tail will likely require further assumptions or different interaction mechanisms beyond those of the unified model framework. Second, even when observed, this behavior is typically confined to an extremely small fraction of the population, for instance, the top $0.1\%$. This is considerably smaller than empirical estimates such as the top $4\%$ reported by \cite{LudwigPhysicsinspiredAnalysisTwoclass2022}.

These limitations highlight several avenues for future research. First, deriving analytical solutions for the stationary distribution is important for definitively characterizing the tail behavior and potentially elucidating the mechanism underlying the inequality reduction at intermediate $\alpha$. Second, incorporating agent heterogeneity based on macroscopic categories, such as skills and age groups, is essential for enhancing the model’s realism. Third, extending the model to more general economic and institutional contexts, for instance, by including elements such as taxation systems and wealth redistribution policies is vital for evaluating its broader applicability and policy implications.

\section*{Acknowledgements}
This work was supported by JSPS KAKENHI Grant Numbers 19K02065, 25K05449.

\clearpage

\bibliographystyle{apalike}
\bibliography{Competition_and_Investment_Model_of_Wealth_Distribution} 

\end{document}